\documentclass[prb,twocolumn,superscriptaddress]{revtex4-1}
\usepackage{graphicx}
\usepackage{amsmath}
\usepackage{hyperref}
\usepackage{color}

\begin{document}

\title{Phenomenological approach to hyperkagome spin liquid:\\gauge fields and spinons without slave particles}

\author{Yuan Wan}
\affiliation{Perimeter Institute for Theoretical Physics, Waterloo, Ontario N2L 2Y5, Canada}
\author{Yong Baek Kim}
\affiliation{Perimeter Institute for Theoretical Physics, Waterloo, Ontario N2L 2Y5, Canada}
\affiliation{Department of Physics, University of Toronto, Toronto, Ontario M5S 1A7, Canada}
\affiliation{Canadian Institute for Advanced Research, Quantum Materials Program, Toronto, Ontario MSG 1Z8, Canada}

\date{\today}

\begin{abstract}
A number of experiments on the hyperkagome iridate, Na$_4$Ir$_3$O$_8$, suggest existence of a gapless quantum spin liquid state at low temperature. Circumventing the slave particle approach commonly used in theoretical analyses of frustrated magnets, we provide a more intuitive, albeit more phenomenological, construction of a quantum spin liquid state for the hyperkagome Heisenberg model.  An effective monomer-dimer model on the hyperkagome lattice is proposed \textit{a la} Hao and Tchernyshyov's approach cultivated from the Husimi cactus model. Employing an arrow representation for the monomer-dimer model, we obtain a compact $U(1)$ gauge theory with a finite density of Fermionic spinons on the hyperoctagon lattice. The resulting theory and its mean field treatments lead to remarkable agreement with previous slave-particle construction of a quantum spin liquid state on the hyperkagome lattice. Our results offer novel insights on theoretical understanding of the emergence of spinon Fermi surfaces and useful predictions for future experiments.
\end{abstract}

\maketitle

\section{Introduction}

\begin{figure}
\includegraphics[width = \columnwidth]{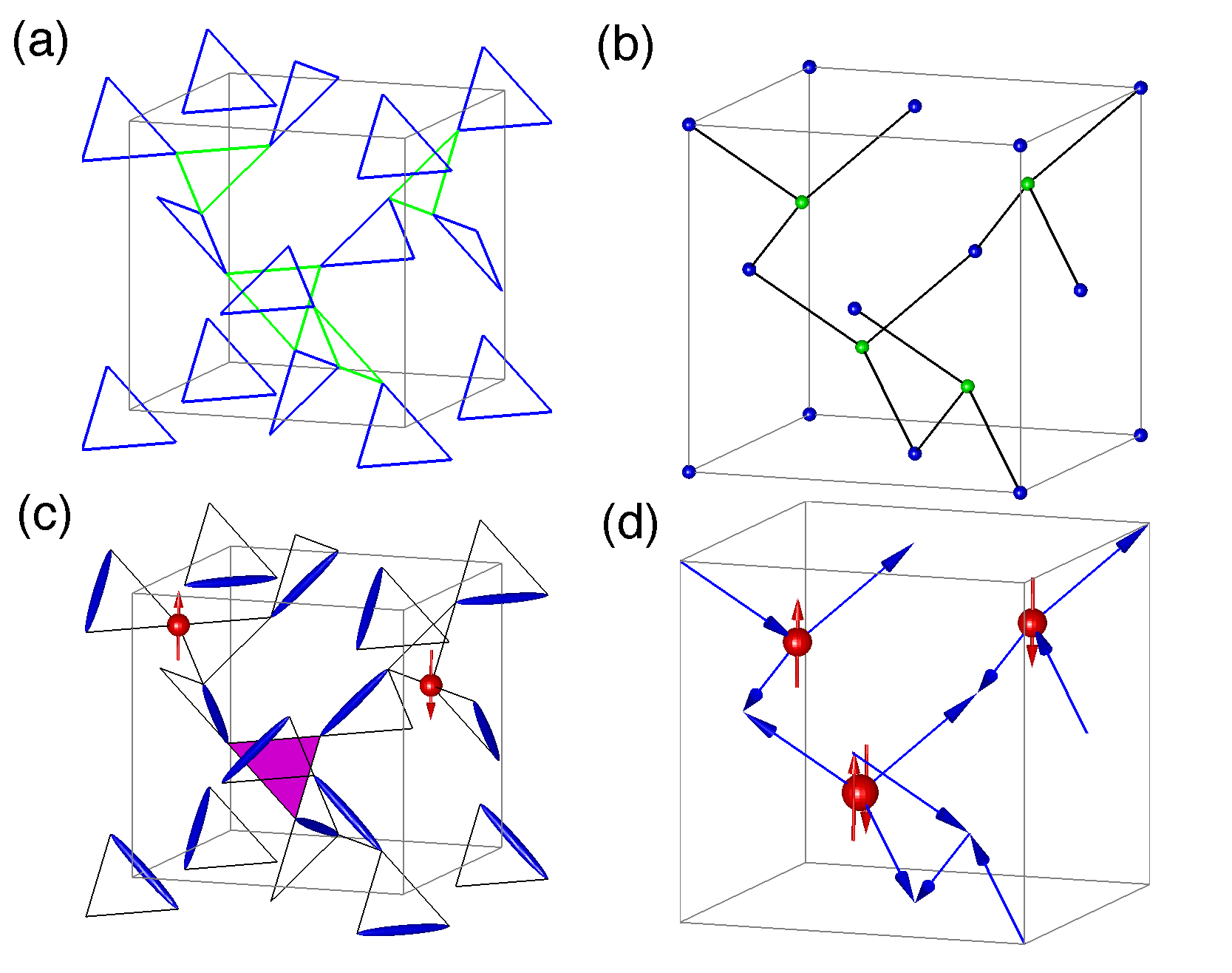}
\caption{(Color online). (a) Hyperkagome lattice. Two sets of interpenetrating triangles are colored in blue and green. Grey box shows a unit cell. (b) Hyperoctagon lattice. A and B sub-lattice sites are colored in blue and green, respectively. Grey box shows a unit cell. The midpoints of hyperoctagon links form a hyperkagome lattice. (c) The monomer-dimer model. Blue ellipsoids represent dimers, \textit{i.e.} the spin singlets formed by nearest-neighbor spins. Red spheres represent spinons or fermionic monomers. Arrows indicate the spin orientation of spinons. A defect triangle is colored in magenta. (d) Arrow configuration on hyperoctagon lattice, which corresponds to the monomer-dimer state presented in (c). Note an empty triangle is regarded as a pair of spinons with opposite spins.}
\label{fig:lattice}
\end{figure}

Hyperkagome iridate, Na$_4$Ir$_3$O$_8$, is a rare candidate material for three-dimensional quantum spin liquid phase, a quantum paramagnet without spontaneous symmetry breaking~\cite{Okamoto2007,Balents2010}. Here the pseudospin $S=1/2$ local moments reside on hyperkagome lattice, a three-dimensional network of corner-sharing triangles (Fig.~\ref{fig:lattice}a). While it is an insulator, the specific heat coefficient $\gamma= C/T$ and uniform susceptibility $\chi$ show the behavior that would be typical for a metal, \textit{i.e.} both remain finite down to the lowest experimentally accessible temperature. This peculiar phenomenology could be explained if the ground state of the material is a quantum spin liquid hosting fermionic excitations known as spinons. Each spinon carries $S=1/2$ but no electric charge. The finite $\gamma$ and $\chi$ at zero temperature is attributed to the Fermi surfaces of spinons~\cite{Lawler2008b,Zhou2008}. The existence of spinon Fermi surfaces is also consistent with recent thermal conductivity and Knight shift measurements~\cite{Singh2013,Shockley2015,Fauque2015}.

On the theory front, it has been argued that the hyperkagome Heisenberg antiferromagnet (KHAHF) is a good first approximation to  Na$_4$Ir$_3$O$_8$~\cite{Hopkinson2007,Lawler2008a,Lawler2008b,Zhou2008,Singh2012}. The model Hamiltonian is given by,
\begin{align}
H = J\sum_{\langle ij\rangle}\mathbf{S}_i \cdot\mathbf{S}_j,
\label{eq:hkhaf}
\end{align}
where $\mathbf{S}_i$ is the $S=1/2$ pseudospin operator on site $i$ of the hyperkagome lattice. The summation runs over all nearest-neighbor pairs. $J>0$ is the exchange constant between nearest-neighbor pseudospins. 

There have been several theoretical attempts to explicitly construct a candidate spin liquid ground state for Eq.~\eqref{eq:hkhaf}. Most of these constructions are based on the slave-particle or parton approach~\cite{Lawler2008a,Lawler2008b,Zhou2008}. In this approach, the physical spin is represented as a bilinear combination of the partons, which can be either fermions or bosons. Among these constructions, the one based on fermionic partons seems to be more satisfactory. A Monte Carlo study on the variational wave functions of fermionic partons found that a quantum spin liquid state with spinon Fermi surfaces and fluctuating $U(1)$ gauge field has the lowest energy~\cite{Lawler2008b}. In particular, the prediction of spinon Fermi surfaces lends support to the aforementioned interpretation of the Na$_4$Ir$_3$O$_8$ phenomenology. While the parton construction looks quite successful in capturing the spin liquid ground state of Eq.~\eqref{eq:hkhaf}, a number of questions remain unanswered in this line of approach. For instance, why is the fermionic parton wave function, rather than the bosonic counterpart, more consistent with the experiments? What physical picture is behind the emergence of the spinon Fermi surfaces?

In this work, we provide an alternative, and more intuitive, construction of the quantum spin liquid state for Eq.~\eqref{eq:hkhaf}. Our approach is motivated by insights advocated by Hao and Tchernyshyov in their studies of the Husimi cactus and the two-dimensional kagome Heisenberg antiferromagnet~\cite{Hao2009,Hao2013}. The basic idea is to use the low energy degrees of freedom of the antiferromagnetic Heisenberg model on Husimi cactus lattice~\cite{Elser1993}, which is a tree analog of the kagome and hyperkagome lattices, as the building blocks for the effective theory of HKHAF. 

In the Husimi cactus model, the ground states are singlet coverings, where every triangle is occupied by exactly one singlet formed by nearest-neighbor spins. The low energy $S=1/2$ excitations are created from such a \emph{vacuum}. As we shall see later in the main text, starting from a defect triangle, which is not occupied by a singlet, a pair of spinons with $S^z=1/2$ and -1/2 can be created by local exchange interactions. The motion of these {\it spinon} excitations occurs with successive flipping of a string of nearby singlets across the lattice. In Husimi cactus model, these spinons are shown to possess fermionic statistics. Thus, the low energy degrees of freedom of the Husimi cactus model consists of nearest-neighbor singlets and ferminoic spinons.

Given that the Husimi cactus and the hyperkagome lattice share the same local structure, we expect that the low-energy degrees of freedom of HKHAF are also nearest-neighbor singlets and fermionic spinons. This observation leads to a monomer(spinon)-dimer(singlet) model on the hyperkagome lattice (Fig.~\ref{fig:lattice}c). It can be shown that there are not enough dimers to cover every triangle on the hyperkagome lattice and a quarter of triangles remain as defect triangles. As a pair of spinons can be generated from each defect triangle, the density of monomers per spin flavor in the monomer-dimer model should be $f=1/4$. 

It is convenient to map the dimer coverings on the hyperkagome lattice to patterns of arrows connecting the centers of triangles. The triangle centers of the hyperkagome lattice formed the so-called hyperoctagon lattice~\cite{Hermanns2014}. Hence, the arrows reside on the links of the hyperoctagon lattice. The spinons are mapped to point defects residing on the hyperoctagon sites (Fig.~\ref{fig:lattice}b,d). The density of these fermionic defects is again 1/4. The resulting model is shown to be a compact lattice $U(1)$ gauge theory, where the fermionic defects or spinons carry the gauge-charge 1. The mean field band structure of spinons at a quarter filling produces multiple Fermi surfaces with one hole-like and two particle-like pockets. Remarkably, this is precisely the same spinon band structure obtained in the previous slave-fermion mean field theory~\cite{Lawler2008b,Zhou2008}. Hence the monomer-dimer model and its arrow representation provide much needed physical insights for the Fermi surfaces of spinons on the hyperkagome lattice. Utilizing this picture, we have also computed the spin structure factor that can be measured in scattering experiments and point out various salient features that arise from the spinon Fermi surfaces.

The rest of the paper is organized as follows. In Section \ref{sec:model}, we construct the phenomenological lattice $U(1)$ gauge theory for the HKHAF. In Section \ref{sec:mft}, we study the $U(1)$ gauge theory at the mean field level and find that the model provides a natural access to both $U(1)$ and $Z_2$ quantum spin liquid ground state. As a simple application of our theory, we also compute the dynamical spin structure factor in the $U(1)$ spin liquid state of the HKHAF. In Section \ref{sec:discuss}, we discuss potentially interesting directions to explore in relation to experiments on hyperkagome spin liquid materials such as Na$_4$Ir$_3$O$_8$.

\section{Model}
\label{sec:model}

In this section, we construct the phenomenological lattice gauge theory for HKHAF. We follow the previous construction for the kagome Heisenberg antiferromagnet in Ref.~[\onlinecite{Hao2013}]. For the sake of being self-contained, in Sec.~\ref{subsec:husimi}, we briefly review the results about the $S=1/2$ antiferromagnetic Heisenberg model on Husimi cactus. Based on these results, we motivate a monomer-dimer model in hyperkagome lattice in Sec.~\ref{subsec:monomer-dimer}. In Sec.~\ref{subsec:arrow},  by using the arrow representation~\cite{Elser1993,Hao2013}, we map the Hilbert space of the hyperkagome monomer-dimer model to arrow configurations on the links of the hyperoctagon lattice, which is the key step toward the lattice gauge theory. Finally, we write down the phenomenological lattice gauge theory Hamiltonian in Sec.~\ref{subsec:gauge}.

\subsection{\label{subsec:husimi}Husimi cactus model}

\begin{figure}
\includegraphics[width = \columnwidth]{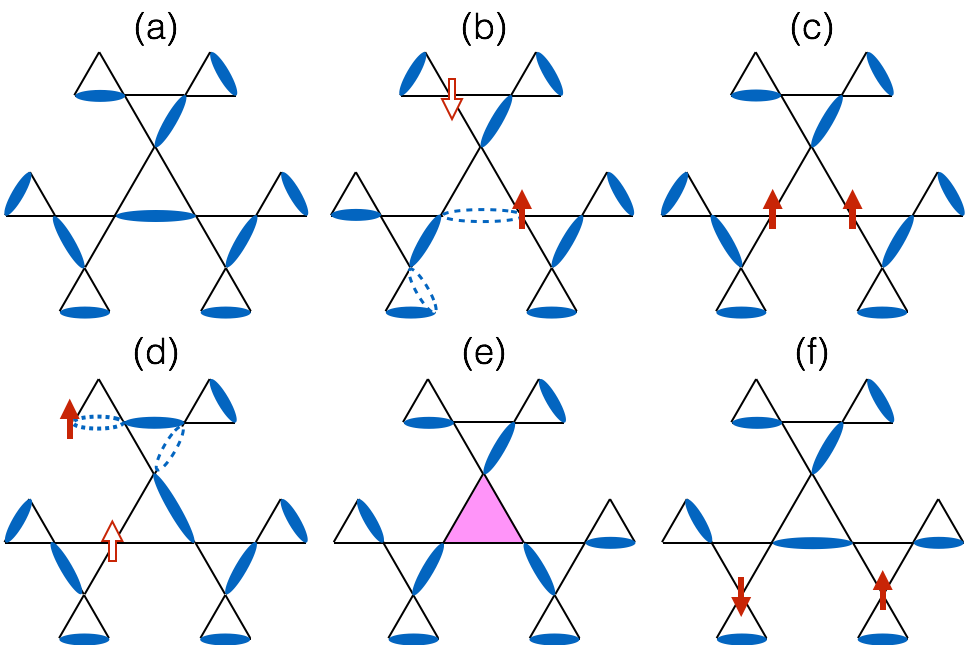}
\caption{(Color online). (a) A ground state of the Husimi cactus model. Each ellipse represents a singlet. (b) A spinon (red solid uparrow) and an anti-spinon (red empty downarrow). The spinon may propagate in the Husimi cactus by flipping a string of singlets. Empty ellipses show the resulted singlet covering pattern after the spinon moves along a path. The anti-spinon is immobile. (c) Breaking up a singlet in the ground state given in (a) creates two unpaired spins. (d) One of the two spins in (c) may hop away by permuting singlets along a path, which becomes a spinon (red solid arrow), while the other becomes an anti-spinon (red empty arrow). (e)(f) A defect triangle (magenta shaded triangle) that is neither covered by a singlet nor occupied by spinons breaks up into a pair of spinons (red solid arrows) under the action of local exchange interaction.}
\label{fig:husimi}
\end{figure}

In this section, we briefly review the results of the antiferromagnetic Heisenberg model on Husimi cactus~\cite{Elser1993,Hao2009}. These results later will be used to motivate the hyperkagome monomer-dimer model.

The Husimi cactus is a tree network of corner-sharing triangles (Fig.~\ref{fig:husimi}a). The antiferromagnetic Heisenberg model Hamiltonian is given by,
\begin{align}
H_\mathrm{Husimi}  =  J\sum_{\langle ij\rangle}\mathbf{S}_i\cdot\mathbf{S}_j = \frac{J}{2}\sum_{\bigtriangleup} \mathbf{S}^2_\bigtriangleup + \mathrm{const}.
\label{eq:husimi}
\end{align}
Here, $\mathbf{S}_i$ is the $S=1/2$ spin operator on cactus site $i$. The exchange constant $J>0$. The first summation is over all nearest neighbour pairs in the cactus, whereas the second summation is over all triangles. The total spin in a given triangle $\mathbf{S}_\bigtriangleup = \sum_{i\in\bigtriangleup}\mathbf{S}_i$, where the summation is over all three sites belonging to the triangle.

The energy of Eq.~\eqref{eq:husimi} is minimized if $\mathbf{S}_\bigtriangleup=1/2$ for every triangle. This can be attained by putting nearest-neighbor spins in spin singlets and covering every triangle by a singlet (Fig.\ref{fig:husimi}a). Since the cactus can be covered in numerous ways, Eq.~\eqref{eq:husimi} has massively degenerate ground states. A triangle that is covered by a singlet is known as a ``vacuum'' triangle since the local interaction energy is minimized.

Eq.~\eqref{eq:husimi} supports $S=1/2$ excitations, which are unpaired spins (Fig.~\ref{fig:husimi}b)~\cite{Hao2009}. It is necessary to distinguish two types of $S=1/2$ excitations, namely spinons and anti-spinons.  A spinon is sandwiched by a vacuum triangle and a triangle that is not covered by a singlet, whereas an anti-spinon is sandwiched by two vacuum triangles. A spinon may propagate in the cactus under the action of local exchange interaction. By contrast, an anti-spinon is immobile. In fact, an anti-spinon is an eigenstate of Eq.~\eqref{eq:husimi}, whose energy is accidentally degenerate with the aforementioned ground states. Furthermore, a braiding argument shows that both spinons and anti-spinons carry fermion statistics. Finally, breaking up a singlet creates a spinon and an anti-spinon from the ground state, and hence the nomenclature (Fig.~\ref{fig:husimi}c,d).

The cactus also supports non-magnetic excitations. To see this, we consider a state in which all but one triangles are covered by singlets. In Fig.\ref{fig:husimi}e, the central triangle (magenta) is neither covered by a singlet nor an unpaired spin. Such a triangle is known as a ``defect'' triangle. Applying the Hamiltonian Eq.~\eqref{eq:husimi} to the defect triangle creates two spinons, which may propagate independently under the action of the Hamiltonian (Fig.\ref{fig:husimi}f). Calculations show that a defect triangle should be regarded as a $S=0$ bound state of two $S=1/2$ spinons, which indicates that there is an attraction potential between two spinons in the Husimi cactus model.~\cite{Hao2009}

\subsection{\label{subsec:monomer-dimer} Monomer-Dimer model}

Both kagome and Husimi cactus feature the motif of corner-sharing triangles. It is therefore reasonable to expect that the local physics of the kagome Heisenberg antiferromagnet is similar to that of the Husimi cactus model. In particular, one may use the low-energy, local degrees of freedom in the Husimi cactus model to motivate a kagome monomer-dimer model~\cite{Hao2013}. Note that the resulting model necessarily contains physics beyond the Husimi cactus as the kagome lattice and the Husimi cactus differ globally; the former contains loops of triangles whereas the latter has none.

Given that the hyperkagome shares the same motif, we may adapt the aforementioned methodology to HKHAF and motivate a similar monomer-dimer model (Fig.~\ref{fig:lattice}c). We surmise that this approach will work even better for the HKHAF because the hyperkagome lattice is closer to the Husimi cactus in terms of lattice topology than the kagome lattice. The length of the shortest loop in the hyperkagome is 10, whereas the minimal loop length is 6 in the kagome lattice.

To begin with, we assume that the singlet coverings on hyperkagome belong to the low energy Hilbert space of the Eq.~\eqref{eq:hkhaf}. In the same spirit as the quantum dimer model, we treat two different singlet coverings as being orthogonal to each other. Thus, the spin singlets become dimers. 

In analogy with the Husimi cactus model, we would like to cover each triangle with a dimer (singlet). Had we been able to do so, the resulted state would be an exact ground state of the Hamiltonian Eq.~\eqref{eq:hkhaf}. However, a simple counting shows that this is impossible. On one hand, we can make $N/2$ dimers from a system of $N$ spins. On the other hand, the hyperkagome lattice with periodic boundary condition contains $2N/3$ triangles. In other words, there are not enough dimers to cover every triangle. Specifically, there are must be a fraction of
\begin{align}
f = 1-\frac{N/2}{2N/3} =\frac{1}{4},
\end{align}
triangles being defect triangles. As discussed in Sec.~\ref{subsec:husimi}, each defect triangle can be regarded as a $S=0$ pair of two spinons, which may separate under the action of local exchange interaction. We therefore must include $S=1/2$ spinons in the low energy Hilbert space as well. By contrast, we exclude anti-spinons because we must break a singlet to create an anti-spinon, which is a process costing the energy of order $J$. Furthermore, we endow spinons with Fermion statistics as the case for Husimi cactus model. In the language of quantum dimer model, the spinons are fermonic monomers carrying $S=1/2$ spin. The density of monomers per spin flavor is $f=1/4$.

\subsection{\label{subsec:arrow}Arrow representations of low energy states}

\begin{figure}
\includegraphics[width = \columnwidth]{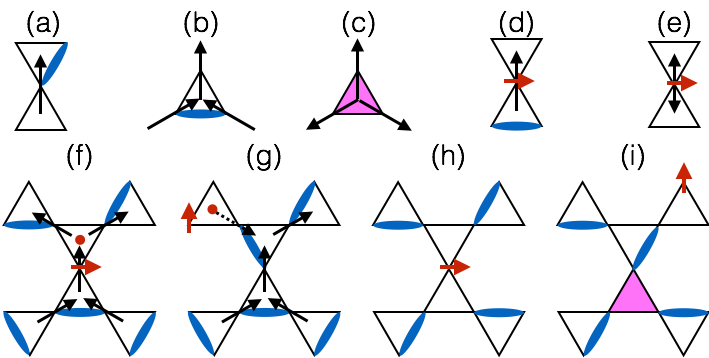}
\caption{(Color online). (a) Arrow rule for a site touched by a dimer. The arrow points in the same direction as the dimer. (b) A vacuum triangle is associated with two incoming and one out-going arrows. (c) A defect triangle is associated with three out-going arrows. (d) Arrow rule for a site occupied by a spinon. The arrow points from the vacuum triangle to the other triangle. (e) A case in which the arrow representation becomes ambiguous. The arrow can point in either directions. (f) A spinon (red short arrow) can be regarded as a point defect (red dot) associated with two out-going arrows and one in-coming arrow. (g) A spinon may hop by flipping a neighboring dimer. In the arrow representation, this corresponds to flipping an arrow that points toward the point defect (black dashed arrow).(h,i) The object shown in (e) is in fact a cluster of three spinons. Hopping the spinon away results in a spinon and a defect triangle (magenta shaded triangle), which is a bound pair of spinons.}
\label{fig:arrow}
\end{figure}

In this section, we employ arrow representation to map the states in the hyperkagome monomer-dimer model to arrow configurations in hyperoctagon lattice.~\cite{Zeng1995, Hao2013} The hyperoctagon lattice is made from the standard diamond lattice by selectively deleting $1/4$ links (Fig.\ref{fig:lattice}b)~\cite{Hermanns2014}. Each unit cell contains 8 sites. The midpoints of the hyperoctagon links form a hyperkagome lattice. Furthermore, each site in the hyperoctagon corresponds to the centre of a triangle in the hyperoctagon. The hyperoctagon lattice inherits the bipartite property from the diamond lattice. The two interpenetrating sublattices of the hyperoctagon lattice correspond to the array of ``up'' and ``down'' triangles in hyperkagome. Note that our choice for the hyperoctagon lattice site coordinates is slightly different from the choice made in Ref.~\onlinecite{Hermanns2014}, but the two are topologically equivalent. In what follows, we describe the arrow representation by enumerating all the rules.

\emph{Rule I.} We consider a site that covered by a dimer (Fig.\ref{fig:arrow}a). The site is shared by two triangles, say 1 and 2. We note that the said dimer must cover one of the two triangles, say 1. We then draw an arrow pointing from the center of the triangle 2 to the center of triangle 1. A necessary consequence of this rule is that a vacuum triangle is associated with two arrows pointing towards its center and one arrow pointing away (Fig.\ref{fig:arrow}b). By contrast, a defect triangle is associated with three arrows pointing away from its center (Fig.\ref{fig:arrow}c).

\emph{Rule II.} We then consider a site occupied by a spinon (Fig.\ref{fig:arrow}d). In this case, the site is shared by a vacuum triangle and a triangle that is not covered by singlet. We then draw an arrow that points from the center of the vacuum triangle to the center of the other triangle. Combing it with the Rule I, we see that the vacuum triangle is associated with two in-coming arrows and one out-going arrow as it should be. The other triangle, however, is associated with two incoming arrows and one out-going arrow. It is then natural to regard the spinon as a point defect residing on the centre of the said triangle (Fig.\ref{fig:arrow}f). In the ensuing discussion, we shall refer the point defect as spinons and use the term ``spinon'' and ``point defect'' interchangably.

\emph{Rule III.} We consider a special case in which the spinon is sandwiched by two triangles, none of them being covered by singlets (Fig.\ref{fig:arrow}e). In this case, we may still draw an arrow pointing from one triangle to the other but the direction is ambiguous. As a result, the arrow representation is no longer a one-to-one mapping. We argue that this ambiguity is insignificant as such cases are rare. In fact, the object presented in Fig.~\ref{fig:arrow}e is a cluster of three spinons (Fig.\ref{fig:arrow}h). To see this, we hop the spinon away and an empty triangle is left as a result. The latter is equivalent to a spinon pair. Thus, the original object must consist of three spinons. Given that the spinon density per spin flavor is $f=1/4$, such objects occur with a small probability of $f^2 = 1/16$.

Since the center of hyperkagome triangles form a hyperoctagon lattice, the above rules map the states of monomer-dimer model to arrow configurations on the links of the hyperoctagon lattice. Importantly, spinons are mapped to point defects residing on hyperoctagon sites. A spinon can hop in the hyperkagome by flipping neighboring dimers. In the arrow representation, the dimer flipping in the hyperkagome is then conveniently mapped to arrow flipping. Specifically, the point defect may hop to a neighboring hyperoctagon site by flipping an out-going arrow (Fig.\ref{fig:arrow}f,g).

\subsection{\label{subsec:gauge}Lattice $U(1)$ gauge theory}

After the preparatory steps, we are ready to write down the phenomenological lattice gauge theory for HKHAF. We first parametrize the orientation of the arrow on the hyperoctagon link $rr'$ by a pseudo-spin-$1/2$ variable $T^z_{rr'} \equiv -T^z_{r'r} = \pm 1/2$. $T^z_{rr'} = 1/2$ if the spin points from the site $r$ to $r'$, and $T^z_{rr'} = -1/2$ if otherwise. The spin raising and lowering operators $T^{\pm}_{rr'}$ flip the arrow on the link $rr'$. Note $T^{\pm}_{rr'}=T^{\mp}_{r'r}$ in our convention.

We define the spinon (point defect) creation and annihilation operator $c^\dagger_{r\sigma}$ and $c^{\phantom{\dagger}}_{r\sigma}$, where $r$ labels the hyperoctagon sites and $\sigma$ is the spin of the spinon. They obey the standard fermion algebra since we have endowed the spinon with fermion statistics.
\begin{subequations}
The kinetic term of spinons is then given by,
\begin{align}
H_t = -t\sum_{rr',\sigma}c^\dagger_{r'\sigma}T^{-}_{r'r}c^{\phantom{\dagger}}_{r\sigma} ,\label{eq:kinetic}
\end{align}
where the summation is over all \emph{oriented} links $rr'$. $t>0$ is the hopping amplitude of spinon. The presence of the spin raising operator $T^{-}_{r'r}$ reflects the fact that the spinon hops from site $r$ to $r'$ by flipping an arrow points from $r$ to $r'$. Note that Eq.~\eqref{eq:kinetic} is Hermitian thanks to the fact that $T^{\pm}_{rr'} =  T^{\mp}_{r'r}$.

In the Husimi cactus model, a spin up spinon and a spin down spinon may form a bound state, which implies that there is an effective on-site attraction between spinons with opposite spins. We therefore add a potential term:
\begin{align}
H_v = -v\sum_{r}c^\dagger_{r\uparrow}c^\dagger_{r\downarrow}c^{\phantom{\dagger}}_{r\downarrow}c^{\phantom{\dagger}}_{r\uparrow},
\label{eq:potential}
\end{align}
where the summation is over all hyperoctagon sites. $v>0$ is the attraction potential. We take the ratio $v/t$ as an arbitrary parameter. The full model Hamiltonian is then given by,
\begin{align}
H_\mathrm{gauge} = H_t+H_v.
\label{eq:gauge}
\end{align}
\end{subequations}

Since the arrow configurations and the location of spinons are not independent, we must enforce the following constraint:
\begin{align}
\sum_{r'}T^z_{rr'} = \sum_{\sigma}c^\dagger_{r\sigma}c^{\phantom{\dagger}}_{r\sigma}-1/2.
\label{eq:gauss_law}
\end{align}
The left hand side is proportional to the number of arrows pointing away from the site $r$ minus the number of arrows pointing toward $r$. The right hand side is related to the number of spinons (point defects) on site $r$. Specifically, in the absence of point defect, there are 2 incoming arrows and 1 outgoing arrow. The left hand side is -1/2, which agrees with the right hand side ($0-1/2$). Furthermore, when the site $r$ is occupied by a spinon, there are two out-going arrows and one-incoming arrow, and hence the left hand side is $1/2$, which equals to the right hand side ($1-1/2$). Finally, when there are two spinons on site $r$, which corresponds to a defect triangle, there arrows point away from $r$. The left hand side gives $3/2$, which again agrees with the right hand side ($2-1/2=3/2$). Importantly, the Hamiltonian Eq.~\eqref{eq:gauge} preserves the constraint Eq.~\eqref{eq:gauss_law}.

The above model is a lattice $U(1)$ gauge theory in disguise. To see this, we interpret $T^z_{rr'}$ as the electric fluxes of a $U(1)$ lattice gauge theory. The constraint Eq.~\eqref{eq:gauss_law} thus becomes the usual Gauss law constraint. In particular, a vacuum hyperoctagon site (a vacuum triangle in hyperkagome lattice) carries charge $-1/2$, whereas a spinon carries charge $1$. The system is overall charge neutral as the total charge density is $-1/2+2f = 0$.

We observe that the Hamiltonian Eq.~\eqref{eq:gauge} possesses a (time-independent) $U(1)$ gauge symmetry: $c^\dagger_r \to c^\dagger_r \exp(i\theta)$ and $T^{-}_{r'r} \to T^{-}_{r'r} \exp[i(\theta_r'-\theta_r)]$. Hence, $T^{-}_{rr'}$ may be interpreted as the link variable, and then the kinetic term Eq.~\eqref{eq:kinetic} takes the form of minimal coupling. Furthermore, the Gauss law Eq.~\eqref{eq:gauss_law} is the generator of such gauge transformation.

To sum up, we have constructed a phenomenological lattice $U(1)$ gauge theory for HKHAF. The $U(1)$ gauge field is coupled to charge-1 fermion at filling factor $f=1/4$. Different from a conventional lattice $U(1)$ gauge theory, the electric fluxes are restricted to value $\pm 1/2$, which implies that the theory is both \emph{strongly coupled} and \emph{frustrated} at the lattice scale.\cite{Hermele2004}

\section{\label{sec:mft}Mean field theory}

In this section, we present a mean-field study of the lattice gauge theory Hamiltonian Eq.~\eqref{eq:gauge}. We first set the on-site attraction between spinons to zero in Sec.~\ref{subsec:u1}. The mean field theory yields a $U(1)$ spin liquid with multiple spinon Fermi surfaces. In Sec.~\ref{subsec:dssf}, we compute the dynamical spin structure within the mean field theory. When the on-site spinon attraction potential is non-zero, the spinon Fermi surfaces are unstable toward a BCS state, which we briefly touch upon in Sec.~\ref{subsec:z2}.

\subsection{\label{subsec:u1}$U(1)$ spin liquid}

\begin{figure}
\includegraphics[width = \columnwidth]{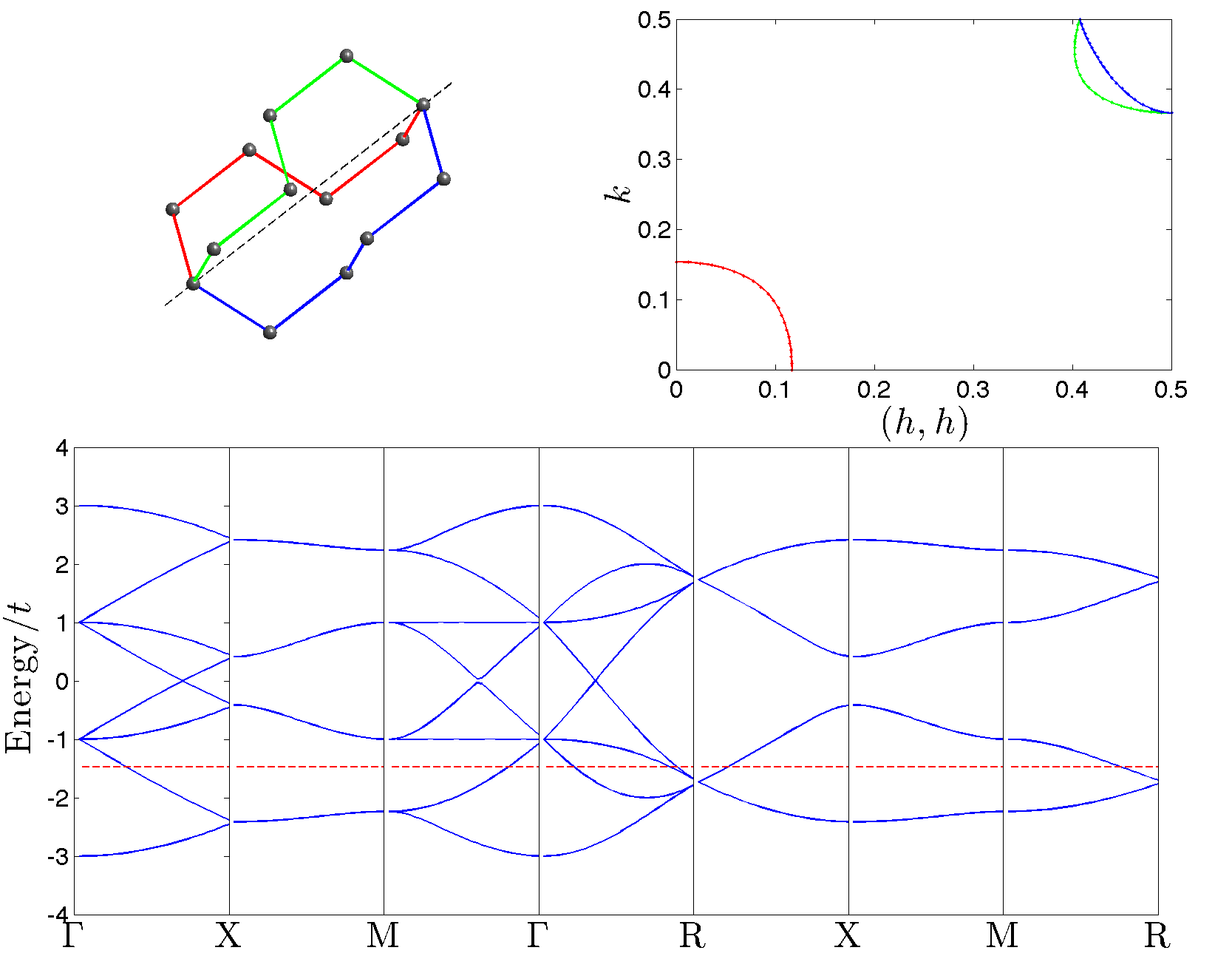}
\caption{(Color online). Top left: A closed surface in the hyperoctagon lattice. Its three edges, colored in red, green, blue respectively, are related by a 3-fold rotational symmetry. The rotational axis is shown as the dashed line. Two edges form a 10-site loop. The three 10-site loops are related by the 3-fold rotational symmetry as well.  Top right: the cross section of the fermi surfaces on the $(h,h,k)$ plane. The hole pocket (red dotted-line) is around $(0,0,0)$ or the $\Gamma$ point. Two particle pockets (green and blue dotted-lines) enclose $(1/2,1/2,1/2)$ or the $\mathrm{R}$ point. Bottom: The spinon dispersion relation along high symmetry directions. Red dashed line shows the position of the Fermi energy.}
\label{fig:tight_binding}
\end{figure}

We first consider the limit $v/t=0$. The kinetic term Eq.~\eqref{eq:kinetic} favors the pseudo-spins to order in the $xy$ plane. Thus, $\langle T^{-}_{r'r}\rangle  \neq 0$. If we assume that the gauge fluctuations are weak, we can simply replace $T^{-}_{r'r}$ by its expectation value, $T^{-}_{r'r} \to \langle T^{-}_{r'r}\rangle$, which is a $U(1)$ phase. In addition, we may replace both sides the Gauss law constraint, Eq.~\eqref{eq:gauss_law}, by its average value, namely $\langle c^\dagger_{r\sigma} c^{\phantom\dagger}_{r\sigma}\rangle =1/4$.  If the time-reversal symmetry is preserved, the flux threading each 10-site loop of the hyperoctagon lattice is either 0 or $\pi$. We consider the simple case in which the fluxes do not break the symmetry of the lattice. In this case, one can show that all fluxes must be 0 by using an argument similar to Ref.~\onlinecite{Zhou2008}. We note three 10-site loops in the hyperoctagon lattice form a closed surface (Fig.~\ref{fig:tight_binding}). On one hand, since these three loops are related by a 3-fold rotational symmetry, the fluxes threading these loops must be equal, say $\Phi$. On the other hand, the total flux going through a closed surface must be $0$ modulo $2\pi$. We thus find $3\Phi = 0$ modulo $2\pi$. Solving this equation, we obtain $\Phi = 0$ modulo $2\pi$.

Since all fluxes are 0, we then further set $\langle T^{-}\rangle_{r'r}  = 1$ for any $r'r$:
\begin{align}
H_\mathrm{gauge} \approx -t\sum_{rr',\sigma}c^\dagger_{r'\sigma} c^{\phantom{\dagger}}_{r\sigma}.
\label{eq:mf_hamil}
\end{align}
The above is simply a free fermion Hamiltonian, which can be readily diagonalized. The spinon band structure along high symmetry directions are shown in Fig.\ref{fig:tight_binding}. We find that, at a quarter filling, the spinons possess three small Fermi pockets: one hole-like pocket surrounding $(0,0,0)$ or the $\Gamma$ point, and two particle-like pockets surrounding $(1/2,1/2,1/2)$ or the R point. Thus, our mean field treatment finds a $U(1)$ spin liquid with spinon Fermi surfaces. Our results are also in a remarkable agreement with the previous studies based on the slave fermion mean field theory, where a $U(1)$ spin liquid state with similar Fermi surface topology was found.~\cite{Lawler2008b,Zhou2008} Here, the result is obtained in a simpler, more intuitive manner from the phenomenological lattice gauge theory.

\subsection{\label{subsec:dssf} Dynamical spin structure factor}

\begin{figure}
\includegraphics[width=0.8\columnwidth]{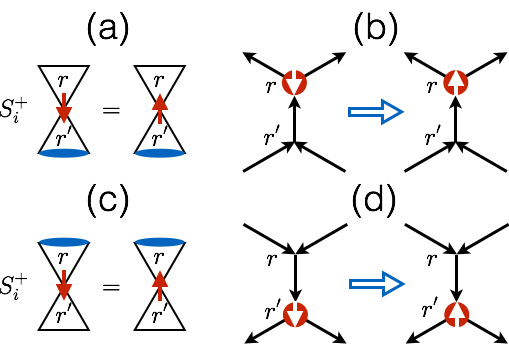}
\caption{(Color online). $S^+_{i}$ raises the spin on hyperkagome site $i$ by 1. Such a process is a low-energy one if and only if a spinon resides on $i$, as shown in (a) and (c). The arrow representation for these processes are presented in (b) and (d).}
\label{fig:spin_operator}
\end{figure}

As a simple application of our phenomenological gauge theory, we compute the zero-temperature dynamical spin structure factor (DSSF) of the $U(1)$ spin liquid state within the mean field approximation. The DSSF is defined as,
\begin{subequations}
\begin{align}
S(\mathbf{q},\omega) \equiv \int dt e^{i\omega t} \langle S^-(\mathbf{q},t) S^+(\mathbf{q},0)\rangle,
\label{eq:dssf}
\end{align}
where,
\begin{align}
S^+(\mathbf{q}) \equiv \frac{1}{\sqrt{N}}\sum_{i}S^+_{i}e^{i\mathbf{q}\cdot\mathbf{x}_i}.
\label{eq:sofq}
\end{align}
\end{subequations}
Here, the summation is over all hyperkagome sites $i$, and $\mathbf{x}_i$ is the position of $i$. $S^\pm_i$ are the spin raising and lowering operators on hyperkagome site $i$. $N$ is the number of hyperkagome sites.

The starting point of our calculation is an explicit relation between $S^\pm_{i}$, which act on the microscopic spin degrees of freedom, and the operators arising in the lattice gauge theory. $S^\pm_{i}$ flips the spin on hyperkagome site $i$. We note that this is a low-energy process if and only if a spinon resides on $i$; otherwise, $S^\pm_{i}$ would break a spin singlet and cost energy of order $J$. Hence, we need only consider the states in which $i$ is occupied by a spinon. Recall that a spinon is sandwiched by a triangle covered by a singlet and a triangle that is not.  Thus, there are two cases to be considered, shown in Fig.~\ref{fig:spin_operator}a \& c, which correspond to the singlet covering the top (labeled $r$) or bottom (labeled $r'$) triangles, respectively.

We first consider the case in Fig.~\ref{fig:spin_operator}a. In terms of arrow representation, the spinon or point defect resides on the hyperhoneycmob lattice site $r$. The arrow on link $rr'$ points from $r'$ to $r$ (Fig.~\ref{fig:spin_operator}b). Thus, $T^z_{rr'}=-1/2$. Note the arrows or $T^z_{rr'}$ variables remain the same after the spin flip. Likewise, the arrow representation for Fig.~\ref{fig:spin_operator}c is shown in Fig.~\ref{fig:spin_operator}d. In this case, the spinon or point defect resides on the hyperoctagon site $r$, and $T^z_{rr'}=1/2$. 

Taking both cases into account, we postulate the following relation:
\begin{subequations}
\begin{align}
S^+_{i} \approx c^\dagger_{r\uparrow}c^{\phantom\dagger}_{r\downarrow} T^-_{rr'}T^+_{rr'}+c^\dagger_{r'\uparrow}c^{\phantom\dagger}_{r'\downarrow} T^+_{rr'}T^-_{rr'}.
\label{eq:spin_op}
\end{align}
The first term on the right hand side of Eq.~\eqref{eq:spin_op} corresponds to the case shown in Fig.~\ref{fig:spin_operator}a. $c^\dagger_{r\uparrow}c^{\phantom\dagger}_{r\downarrow}$ flips the spin of the spinon (point defect) on hyperoctagon site $r$. $T^-_{rr'}T^+_{rr'}$ projects onto the subspace with $T^z_{rr'}=-1/2$. Likewise, the second term corresponds to the case shown in Fig.~\ref{fig:spin_operator}. Note Eq.~\eqref{eq:spin_op} is symmetric when exchanging $r$ and $r'$.

Within the mean field theory, we may further neglect the fluctuation of $T^z_{rr'}$, and replace $T^+_{rr'}T^-_{rr'}$ and $T^-_{rr'}T^+_{rr'}$ by their classical expectation value $1/2$. We thereby obtain a simple relation between $S^+_{i}$ and the spinon operators:
\begin{align}
S^+_{i} \approx (c^\dagger_{r\uparrow}c^{\phantom\dagger}_{r\downarrow}+c^\dagger_{r'\uparrow}c^{\phantom\dagger}_{r'\downarrow})/2.
\label{eq:spin_op_mf}
\end{align}
\end{subequations}

Plugging Eq.~\eqref{eq:spin_op_mf} into Eq.~\eqref{eq:sofq}, we find,
\begin{align}
S^+(\mathbf{q}) \approx \sum_{\mathbf{k},\alpha} f_{\alpha}(\mathbf{q})c^\dagger_{\alpha\uparrow}(\mathbf{k}+\mathbf{q})c^{\phantom\dagger}_{\alpha\downarrow}(\mathbf{k}).
\end{align}
We have omitted an over-all normalization constant. $\alpha$ runs over 8 sublattice labels of the hyperoctagon lattice. $c_{\alpha\sigma}(\mathbf{k})$ destroys a spinon (point defect) with momentum $\mathbf{k}$, sublattice index $\alpha$, and spin $\sigma$. The form factor $f_\alpha(\mathbf{q}) = \sum_{\mathbf{d}_\alpha}\exp(i\mathbf{q}\cdot\mathbf{d}_\alpha/2)$, where the summation is over all vectors $\mathbf{d}_\alpha$ pointing from the hyperhoneycmob sublattice site $\alpha$ to its nearest-neighbors. The factor of $1/2$ in the exponential comes from the fact that the hyperkagome sites are the midpoints of the hyperoctagon links.

\begin{figure}
\includegraphics[width=\columnwidth]{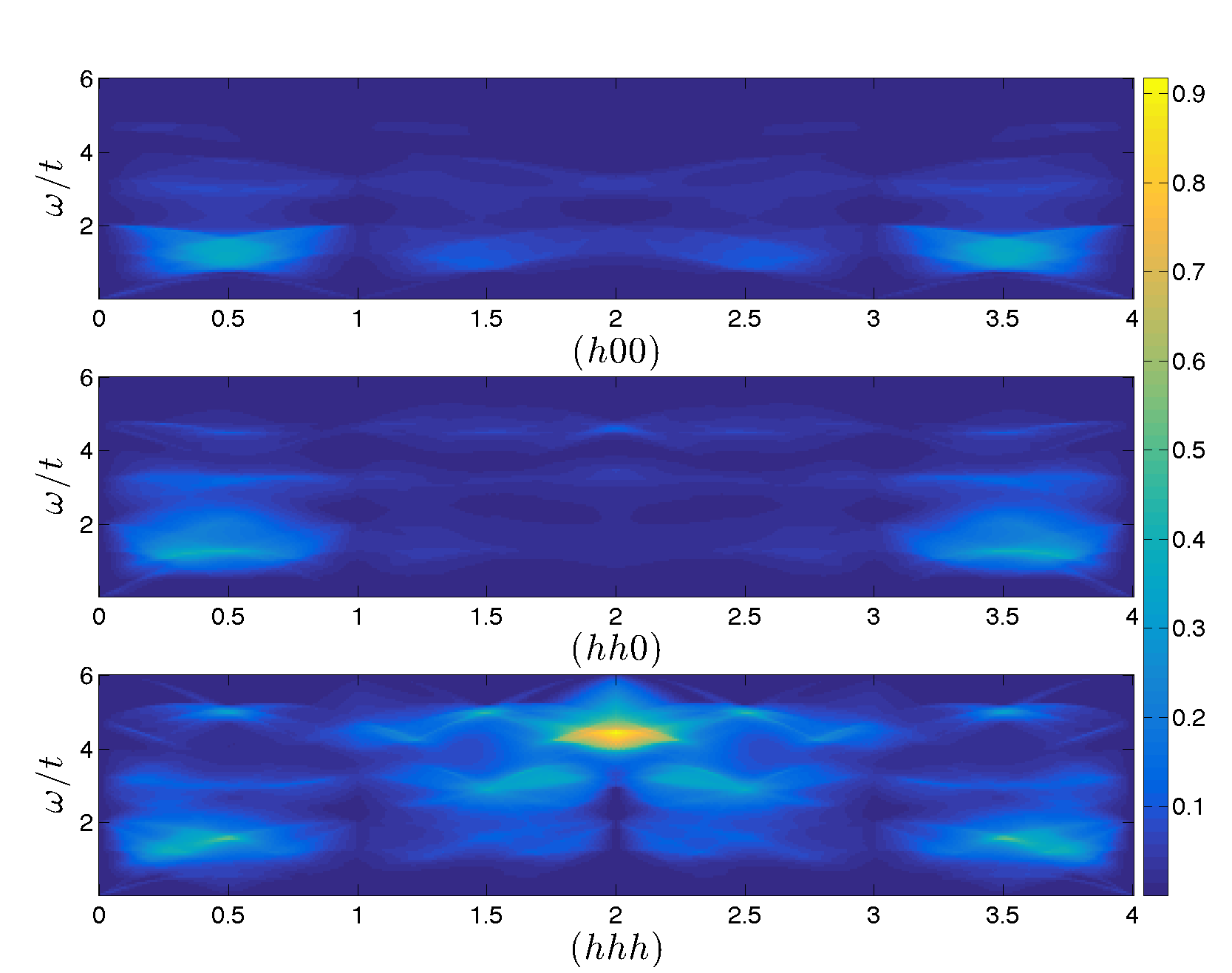}
\caption{(Color online). Dynamical spin structure factor $S(\mathbf{q},\omega)$ along three high symmetry directions. The same color scale is used for all three plots.}
\label{fig:sqw_0to6}
\end{figure}

\begin{figure}
\includegraphics[width=\columnwidth]{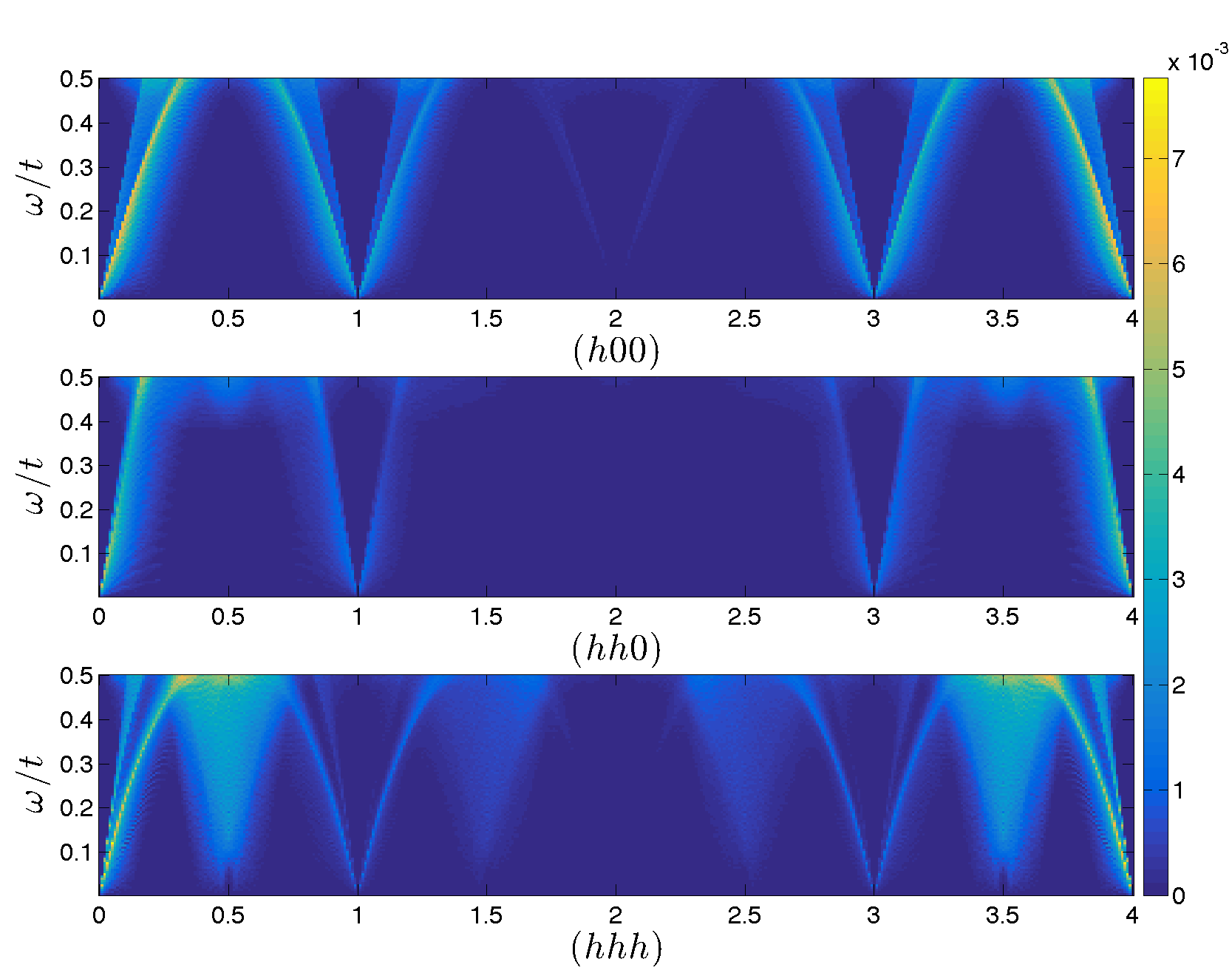}
\caption{(Color online). The same as Fig.~\ref{fig:sqw_0to6} but for a smaller energy window $[0,0.5]t$.}
\label{fig:sqw_0to0p5}
\end{figure}

Within the mean field theory, the problem of calculating the DSSF Eq.~\eqref{eq:dssf} in the $U(1)$ spin liquid phase is reduced to the standard problem of computing two-body correlation function of the free-fermion Hamiltonian Eq.~\eqref{eq:mf_hamil}. The results are presented in Fig.~\ref{fig:sqw_0to6}. Note that, within the mean field theory, the DSSF vanishes when $\omega$ is greater than the bandwidth of the spinon, which is $6t$. We find the spectral weight of the DSSF is broadly distributed in both momentum and energy with no sharp features, which is typical of spin liquid. Furthermore, the DSSF has a larger period in the momentum space than the spinon dispersion relation due to the form factor $f_\alpha(\mathbf{q})$. Finally, we observe a strong resonance-like feature near $\mathbf{q} = (2,2,2)$ and $\omega\sim 4t$, and a few weaker features near high symmetry momentum points such as $(0,0,0)$, $(1/2,1/2,0)$, and $(1/2,1/2,1/2)$.

However, we must caution that the above result is based on a simple mean field theory calculation. We don't expect all features will be stable against gauge fluctuations. Yet, we believe that the low-energy features of the DSSF are likely to be more robust. To this end, we zoom in to the lower energy window, $\omega \in [0,0.5]t$ (Fig.~\ref{fig:sqw_0to0p5}). Note the overall spectral weight is significantly smaller as shown in the color bar. We find that there are dispersive, sharp resonance-like features emanating from the $\mathbf{q}=0$, which we attribute to the intra-pocket particle-hole transitions and the transition between the two pockets around $\mathrm{R}$ point. In addition, the quasi-nesting between the Fermi pockets near $\Gamma$ and $\mathrm{R}$ points produces another resonance-like feature near the nesting vector $\mathbf{q}=(1/2,1/2,1/2)$. Since these low-energy features are directly tied to the topology of the spinon Fermi surface, they will serve as a powerful probe for the underlying spinon Fermi surface in scattering experiments.

\subsection{\label{subsec:z2} Instability toward $Z_2$ spin liquid}

So far we have been focusing on the limit $v/t=0$.  In this limit, the mean field treatment yields a $U(1)$ spin liquid with spinon Fermi surfaces. Therefore, when $v>0$, such a state may be unstable toward a BCS state of spinons, or equivalently a $Z_2$ spin liquid. However, this picture is based on the assumption that the fluctuations are negligible. A more rigorous variational calculation on the HKHAF spin Hamiltonian Eq.~\ref{eq:hkhaf}, in which fluctuations are partially accounted for, shows that the energy of the $U(1)$ spin liquid state is in fact lower than the proximate $Z_2$ spin liquid state.~\cite{Lawler2008b} 

Nevertheless, one may ask which spinon BCS state will be stabilized if the HKHAF model Eq.~\eqref{eq:hkhaf} is perturbed in such a way that a $Z_2$ spin liquid is favored. In this case, a mean field theory calculation similar to the previous subsection finds a simple $s$-wave BCS state with fully gapped spinons. The spinon gap is very small due to the small density of states at the Fermi energy. For $v/t = 1$, we find the spinon gap $\Delta/t=2.747\times10^{-6}$. Given such a small energy scale, it would be very difficult to detect the $Z_2$ spin liquid state experimentally or numerically.

\section{Discussion}
\label{sec:discuss}

In this paper, we have presented a phenomenological lattice $U(1)$ gauge theory for the HKHAF. We have shown that the theory provides access to both $U(1)$ and $Z_2$ spin liquid states in a natural, intuitive manner. Furthermore, it points toward various potentially interesting directions to explore in future.

In a $U(1)$ spin liquid without global spin symmetry such as the quantum spin ice, the photon will contribute a sharp, dispersive resonance in the DSSF, which would be an unambiguous experimental signature for the underlying $U(1)$ spin liquid state~\cite{Hermele2004,Damle_Kim2008,Benton2012}. By contrast, the HKHAF model Hamiltonian Eq.~\eqref{eq:hkhaf} possesses a global pseudospin $SU(2)$ symmetry. The gauge electric field operators $T^z_{rr'}$ transform trivially under global spin rotations. The photon excitation therefore carries $S=0$. As a result, the photon excitation is invisible in the DSSF, which probes $S=1$ excitations. It would be interesting to devise an experimental measure to probe the photon excitation in the KHAHF and predict from theory its experimental signature. We note that, in the context of the kagome $U(1)$ Dirac spin liquid, it has been shown that Dzyaloshinskii-Moriya (DM) interaction, which breaks the global spin $SU(2)$ symmetry down to $U(1)$, may endow the gauge fluctuations with a small magnetic dipole moment and make the photon mode visible in the DSSF~\cite{Lee2013}. Given that the hyperkagome lattice lacks bond inversion symmetry, the DM interaction is present on symmetry ground~\cite{Chen2008}. Different from the kagome Dirac spin liquid, the photon in the hyperkagome spin liquid is likely damped by the soft particle-hole excitations near the spinon Fermi surfaces. However, provided that the density of states is small near the Fermi energy, we expect that the damping is insignificant.

We have shown that the spinon filling factor $f=1/4$ is fixed by the lattice topology. In the absence of magnetic field, the density of spin-up spinon, $f_\uparrow$, and the density of spin-down spinon, $f_\downarrow$, are equal. An external magnetic field induces the imbalance between $f_\uparrow$ and $f_\downarrow$, which in turn changes the spinon Fermi surface topology. Therefore, the external field provides a means to manipulate the spinon Fermi surface. In particular, inspecting the spinon dispersion relation (Fig.~\ref{fig:tight_binding}) shows that the external field may induce a Lifshitz transition of the spinons. The drastic change of spinon Fermi surface topology may manifest itself in specific heat, nuclear magnetic resonance, and scattering experiments. Furthermore, the external magnetic field may induce an internal $U(1)$ gauge flux that is directly coupled to spinons, which could produce the thermal Hall effect~\cite{Katsura2010}.

Throughout this work, we assume that the Heisenberg model Eq.~\eqref{eq:hkhaf} is a good first approximation of Na$_4$Ir$_3$O$_8$. Even though various perturbations to the Heisenberg model, such as the Dzyaloshinskii-Moriya (DM) interaction and symmetric anisotropic exchange interactions, are likely small in magnitude, they may become important at very low temperature and lead to a quasi-static order~\cite{Chen2008,Kimchi2014,Dally2014,Shockley2015,Shindou2016,Mizoguchi2016,Witczak-Krempa2014}. It is therefore important to understand the impact of these perturbations on the $U(1)$ spin liquid state and, in particular, to examine whether they may drive instabilities toward magnetic order or valence bond solid~\cite{Bergholtz2010}. A primitive analysis shows that the DM interaction induces an effective spin-orbital coupling for the spinons. We find that a small value of DM interaction is sufficient to remove one of the two particle pockets near the $R$ point. More work is needed to clarify its implications. In addition, it has been suggested that Na$_4$Ir$_3$O$_8$ is proximate to a metal-insulator transition~\cite{Podolsky2009,Singh2013,Chen2013,Fauque2015}. It would also be interesting to incorporate charge fluctuations to our phenomenological gauge field theory.

\begin{acknowledgments}
This work was supported by Perimeter Institute for Theoretical Physics (YW and YBK), the NSERC of Canada (YBK), and the Center for Quantum Materials at the University of Toronto (YBK). Research at Perimeter Institute is supported by the Government of Canada through the Department of Innovation, Science and Economic Development Canada and by the Province of Ontario through the Ministry of Research, Innovation and Science. This work was also performed in part (YBK) at the Aspen Center for Physics, which is supported by National Science Foundation grant PHY-1066293.
\end{acknowledgments}

\bibliography{hyperkagome}

\end{document}